\documentclass[twocolumn,aps, prl, superscriptaddress, showpacs, amsmath,amssymb]{revtex4}

\usepackage{graphicx}
\usepackage{dcolumn}
\usepackage{bm}

\begin{document}
\title{The Effect of Air on Granular Size Separation in a Vibrated Granular Bed}
\author{Matthias E. M\"{o}bius}
\affiliation{The James Franck Institute and Department of Physics,
The University of Chicago, Chicago, Illinois 60637, USA}
\author{Xiang Cheng}
\affiliation{The James Franck Institute and Department of Physics,
The University of Chicago, Chicago, Illinois 60637, USA}
\author{Peter Eshuis}
\altaffiliation[Permanent address: ]{Physics of Fluids, Physics of
Fluids Group, University of Twente, P.O. Box 217, 7500 AE Enschede,
The Netherlands} \affiliation{The James Franck Institute and
Department of Physics, The University of Chicago, Chicago, Illinois
60637, USA}
\author{Greg~S.~Karczmar}
\affiliation{Department of Radiology, The University of Chicago,
Chicago, Illinois 60637, USA}
\author{Sidney R. Nagel}
\affiliation{The James Franck Institute and Department of Physics,
The University of Chicago, Chicago, Illinois 60637, USA}
\author{Heinrich M. Jaeger}
\affiliation{The James Franck Institute and Department of Physics,
The University of Chicago, Chicago, Illinois 60637, USA}


\date{\today}

\begin{abstract}
Using high-speed video and magnetic resonance imaging (MRI) we study
the motion of a large sphere in a vertically vibrated bed of smaller
grains. As previously reported we find a non-monotonic density
dependence of the rise and sink time of the large sphere. We show
that air drag causes relative motion between the intruder and the
bed during the shaking cycle and is ultimately responsible for the
observed density dependence of the rise time. We investigate in
detail how the motion of the intruder sphere is influenced by size
of the background particles, initial vertical position in the bed,
ambient pressure and convection. We explain our results in the
framework of a simple model and find quantitative agreement in key
aspects with numerical simulations to the model equations.
\end{abstract}

\maketitle

\section{I. Introduction}
Granular size separation is one of the most prominent phenomena in
granular physics \cite{3,31,47}. When a granular mixture is agitated
via an external driving force, grains of different sizes separate
into distinct regions of the container. This behavior sets granular
materials apart from ordinary fluids which typically mix to increase
entropy. A key reason is that granular media are far from thermal
equilibrium, and that therefore the dynamics set up by the driving
force govern their behavior. The phenomenon of granular size
separation is well known. It manifests itself in a wide array of
granular systems such as chute flows, avalanches and rotating drums
\cite{32,48,49,50,51,52}.

Here we concentrate on a vertically vibrated bed in which a large
spherical object, the ``intruder'', is embedded. The excitation
typically causes the intruder to rise to the top of the bed, which
is commonly referred to as the Brazil Nut Effect \cite{2}. However,
the detailed behavior is more complex. In particular the rate of
rising is strongly density dependent and in some experiments light
intruders have been found to sink \cite{16,21}. Efforts to identify
the underlying mechanism for the intruder motion have been stymied
by the need to account for this wide variety of behaviors \cite{43}.
\begin{figure}[t!]
\begin{center}
\includegraphics[width=2.0in]{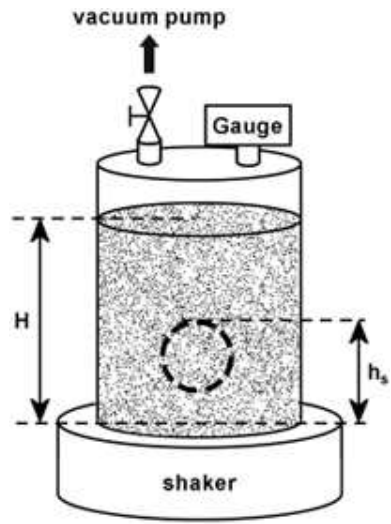}%
\caption{\label{setup} Sketch of the setup. $H$ is the total height
of the bed and $h_s$ the initial height of the intruder measured
from its top. A vacuum gauge measures the background air pressure
inside the cylinder.}
\end{center}
\end{figure}

Until recently, the rising or sinking of an intruder has been
treated as distinct phenomena. Although the interstitial fluid,
typically air, has been shown to strongly influence this density
dependent behavior of the intruder \cite{23,21,41,27}, the details
of how it acts and whether it is the cause of this behavior have
only been resolved recently \cite{29}. While some reports have
suggested that air is unimportant \cite{16,6,30,33,44}, others
acknowledged its role in determining the intruder motion
\cite{21,23,41,42,27,29,26}, but gave conflicting explanations.
There have also been studies of closely related systems with
multiple intruder, quasi 2D and bi-disperse systems
\cite{7,25,22,45,6,26,53}.

This paper presents a systematic study of the effect of air on size
separation in a three-dimensional, single intruder system and
expands on a letter published earlier \cite{29}. In general, size
separation is caused by a variety of mechanisms (for a recent review
see \cite{31}). Our system was chosen such that convection is the
dominant transport mechanism in the absence of air. The separation
mechanism is identified by employing magnetic resonance imaging
(MRI) and high speed video. This allows us to map out in detail the
dependence on system parameters and to establish a phase diagram
that delineates the rising and sinking regimes. We then develop a
model that can explain the key features of our results and also
provides a unifying framework for seemingly conflicting or
unconnected observations made in the literature. In particular, it
shows the sinking and rising to originate from the same mechanism.
We test this model through a simulation.

\section{II. Experimental method}
\begin{figure}
\begin{center}
\includegraphics[width=3.4in]{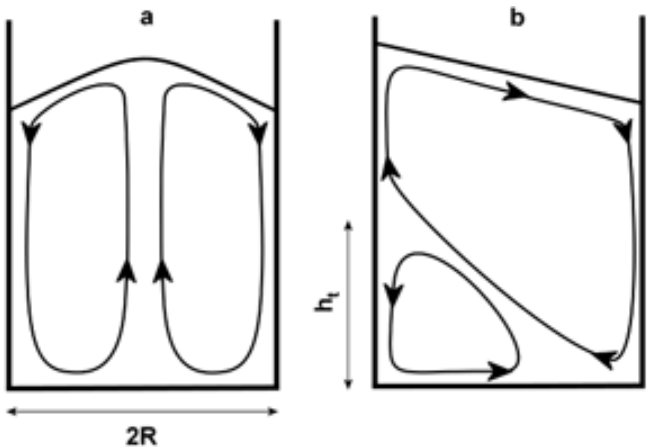}%
\caption{\label{convection} Flow fields of the two observed
convection patterns: (a) axisymmetric, wall-driven convection with a
centered heap surface. Thin downstream at the walls and an upstream
in the middle. (b) Asymmetric, air-driven convection with a slanted
heap surface. Both images are to scale and represent flow patterns
for shaken beds of $d=0.5$ mm glass media. Cell radius is $R=4.1$ cm
and total bed height before shaking is $H=8.5$ cm. The height
$h_t=4.5$ cm separates the regions of up and down flow of the
particles at the wall.}
\end{center}
\end{figure}

All experiments were performed in an acrylic cylinder with an inner
radius of $R=4.1$ cm. It was mounted on a VTS 100 electromagnetic
shaker driven by a function generator that produced sinusoidal
excitations (``taps'') spaced one second apart. Figure \ref{setup}
shows a sketch of the setup. The acceleration was measured in terms
of the dimensionless peak to peak acceleration $\Gamma = a_{p-p} /
2g$ and monitored by an accelerometer (PCB, model 353B03) attached
to the vessel. The frequency of the excitations was typically fixed
at $f=13$ Hz although we explored ranges between 10 and 30 Hz with
no qualitative change in the results. We used different types of
media to study the effect of density: glass ($\rho_m=2.5$ g/ml,
MoSci Corp.), zirconium oxide ($\rho_m=3.8$ g/ml, Glen Mills), DVB
resin ($\rho_m=1.0$ g/ml, Supelco), tapioca pearls ($\rho_m=1.2$
g/ml), poppy and rajagara seeds ($\rho_m=1.0$ and $1.2$ g/ml,
respectively). Here, $\rho_m$ is the density of the bed particle
{\it material}. The range of sizes we used as our bed medium was
between $0.25$ and $2.0$ mm. The intruder particles were $25$ mm
hollow polypropylene spheres (Euro-Matic Plastics Inc.) filled with
varying amounts of material such as lead shot to tune the density.
The intruder was placed along the cylinder axis at height $h_s$
(Fig. \ref{setup}). After each run, the cell was emptied and
refilled to avoid compaction effects.

Two different cells were used: One had a smooth inner wall, the
other was roughened by gluing glass beads to its surface, thus
allowing us to study the effect of wall-driven convection on the
intruder motion \cite{4,37}. During the experiments the cylinder was
closed with a lid that had a pressure gauge and quick release valve
mounted on it. The cell could be evacuated to pressures between
$0.13$ and $101$ kPa. The vacuum pump was then disconnected via the
valve to avoid vibrations caused by dangling tubes.

A thin plastic straw attached to the intruder that extended above
the bed surface enabled us to track the intruder motion inside the
bed. We recorded the motion of the straw with a high-speed video
camera to measure the intruder trajectory during the shaking cycle
as well as the net displacement of the intruder after each tap.
Measurements of the rise time with and without the straw did not
show any difference within experimental accuracy..

Two different techniques were employed to probe the interior of the
three-dimensional particle bed. One of them involved non-invasive
magnetic resonance imaging (MRI). The bed was layered with MRI
active poppy seeds and MRI passive rajagara seeds \cite{29}. As a
result, an MRI image of an axial cut through the bed shows a stack
of bright and dark bands. After each tap and once the system has
settled we take a MRI snapshot, imaging a vertical slice through the
center. From the displacement of the bands between taps we can then
deduce the flow field. While the size and density differences
between the two types of seeds is small, it is not clear a priori
that they do not matter. The poppy seeds are kidney shaped whereas
the rajagara ones are more spherical, although both are of
comparable size ($d\approx 0.7$ mm). In order to ensure that our MRI
results are unaffected by these differences and hold for media other
than seeds, we employed a second technique to visualize the inside
of a bed of glass beads. In this technique, initially the bed was
prepared with equally spaced layers of black and white glass beads.
The flow field at later times was obtained by carefully backfilling
the cell with water, freezing it and then cutting it along the
central axis.
\section{III. Results and Discussion}
{\it Convective flow patterns}. Since convection plays an important
role in granular size separation in systems such as ours \cite{4},
we first investigated the convective flow. There are two types of
convection patterns we observe - one is symmetric and the other
asymmetric (Fig. \ref{convection}). For glass media the symmetric
roll is observed with $d > 0.5$ mm in the smooth cell and $d > 0.35$
mm in the rough cell. This is the well known wall-driven convection
roll \cite{4,37} that has a thin downstream region near the walls
and a large upstream flow in the center of the bed [Fig.
\ref{convection}(a)]. Small background media, in which air effects
are more pronounced due to the lower bed permeability, undergo the
asymmetric convection despite careful leveling procedures. We found
this asymmetric behavior for all glass media with diameters $d\leq
0.5$ mm in the smooth and $d\leq 0.35$ mm in the rough cell. In this
case a surface instability caused by air flow dominates the
wall-driven convection and a large convection roll spanning the
entire diameter of the cell sits on top of a small one [Fig.
\ref{convection}(b)]. The asymmetric roll tends to drive bed
particles towards the wall at which point they either move downward
or upward depending on their vertical position.

{\it Density dependent intruder rise time}. When an intruder is
placed in the bed, it will typically rise upon shaking. For
sufficiently small and light media in which air effects are
important, the rate of rising strongly depends on the relative
density between the intruder and the bed medium $\rho / \rho_m$.
\begin{figure}
\begin{center}
\includegraphics[width=3.4in]{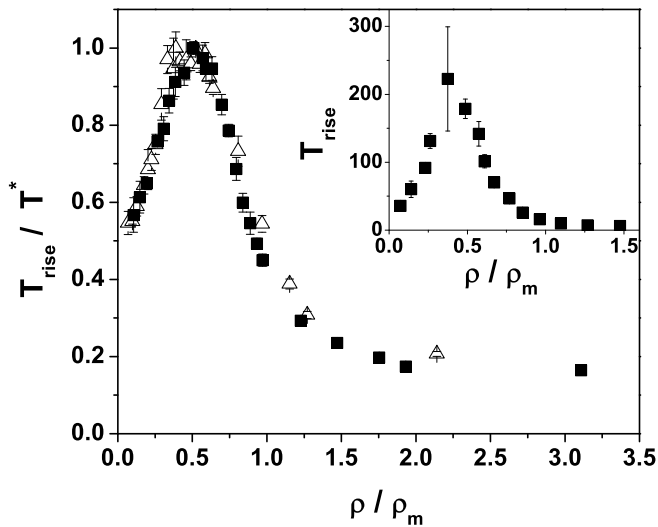}
\caption{\label{diffdensity} Normalized rise time
$T_\textrm{rise}/T^*$ versus $\rho / \rho_m$ for different
background densities. $T^*$ is the rise time at the peak. The size
of the media is $d=0.5$ mm, $h_s=4.5$ cm, $\Gamma=5$ and $f=13$ Hz,
$H=8.5$ cm. Symmetric convection: $(\blacksquare)$, Glass ($\rho_m =
2.5$ g/ml); $(\vartriangle)$, Zirconium Oxide ($\rho_m=3.8$ g/ml) in
rough cell. Inset: asymmetric convection, $(\blacksquare)$, DVB
resin ($\rho_m=1.0$ g/ml) in smooth cell.}
\end{center}
\end{figure}
\begin{figure}
\begin{center}
\includegraphics[width=3.4in]{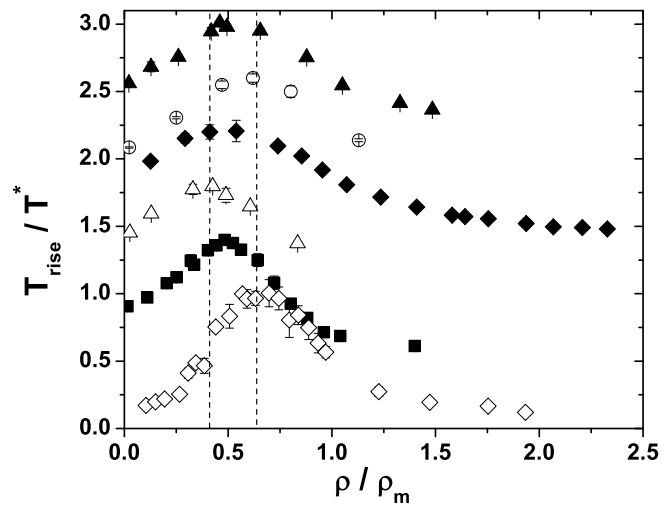}
\caption{\label{diffpar} Normalized rise time $T_\textrm{rise}/T^*$
versus $\rho / \rho_m$ for different excitation parameters, boundary
conditions, interstitial gases and media. In all cases convection is
symmetric and $H=8.5$ cm. For clarity, all the curves except
$(\lozenge)$ have been shifted vertically with separation $0.4$. The
vertical dashed lines are at $\rho/\rho_m=0.41$ and 0.63,
respectively. Unless stated otherwise, we used these standard
parameters: Rough Cell, $R=4.1$ cm, $h_s=4.0$ cm, $d=0.5$ mm glass
beads, $\Gamma=5$ and $f=13$ Hz. $(\blacksquare)$, standard
conditions; $(\blacktriangle)$, $\Gamma=3$; $(\circ)$, $R=6.0$ cm;
$(\blacklozenge)$, poppy seeds $d=0.7$ mm; $(\vartriangle)$, ambient
gas is Helium ($\rho_{\textrm{helium}} \approx
\rho_{\textrm{air}}/7$); $(\lozenge)$, $d=1.0$ mm, smooth cell,
$h_s=4.5$ cm.}
\end{center}
\end{figure}

Despite the two different flow patterns in the bed the behavior of
the intruder is qualitatively the same in both the symmetrical and
asymmetrical patterns as can be seen in Fig. \ref{diffdensity}. The
plot shows the rise time as a function of the intruder density,
where both axes have been rescaled with the convective rise time
$T^*$ and the density of the bed $\rho_m$, respectively. The
non-monotonic curves in Fig. \ref{diffdensity} are typical for
air-driven size separation \cite{23,41,29,27,30}. Light and heavy
intruders rise faster than those at intermediate density.

We define the density at the peak, $\rho^*$, to be the point at
which the rise time $T_\textrm{rise}$ is maximized or diverges as
shown in later plots. In Fig. \ref{diffdensity} $\rho^* / \rho_m
\approx 0.5$. The curves correspond to different background
densities of the same size ($d=0.5$ mm). Both the glass and
zirconium oxide bed exhibit axisymmetric convection as depicted in
Fig. \ref{convection}(a), while the DVB medium does not, despite
having the same size and thus the same permeability. However, due to
its lower density it is more susceptible to air effects which causes
the axisymmetric roll to go unstable and the convective flow becomes
asymmetric as illustrated in Fig. \ref{convection}(b). Instead of
observing a peak at $\rho^{*} / \rho_m$, we find that the rise time
diverges, since the convective flow drives the intruder with $\rho =
\rho^{*}$ to the wall where it stays for long times (sometimes
indefinitely). The inset to Fig. \ref{diffdensity} shows the
divergence at approximately the same relative density as the peak in
the main panel. This means that $\rho^* / \rho_m$ does not depend on
the symmetry of the convective flow.

The ratio $\rho^* / \rho_m$ is almost constant over a range of
system parameters (Fig. \ref{diffpar}). The density of the
interstitial gas and the medium, the size and boundary condition of
the cell and the excitation strength do not significantly influence
the peak position. $\rho^* / \rho_m$ always lies between $0.41$ and
$0.63$. This range of peak positions agrees with recent results by
other groups in similar systems \cite{27,30}. Around the same
relative density Yan et al. \cite{21} furthermore observed a
diverging rise time. However, in their system the intruder sank
below that density.

\begin{figure}
\begin{center}
\includegraphics[width=3.4in]{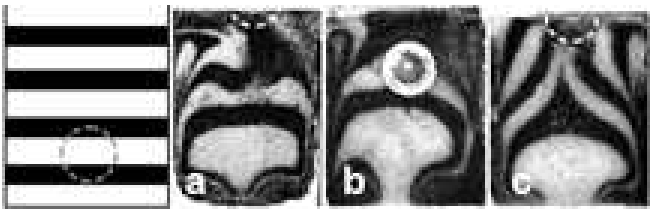}%
\caption{\label{overview} Axial cuts through cell for three
different density intruders. The bed has been cut after it was
saturated with water and frozen. The bed medium consists of $0.5$ mm
glass beads, interleaved with black-colored layers. Left: sketch of
original configuration. (a) $\rho / \rho_m = 0.043$, $26$ taps; (b)
$\rho / \rho_m = 0.5 = \rho^{*} / \rho_m$, $26$ taps; (c) $\rho /
\rho_m = 3.3$, $6$ taps. In (a) and (c) the final intruder positions
are denoted by a dotted circle. Sample (b) was cut with the intruder
still inside the bed.}
\end{center}
\end{figure}

{\it Intruder motion and convective flow}. Our previous MRI
experiments established that intruders with density $\rho^*$ move
with the surrounding bed while light and heavy intruders rise faster
than convection \cite{29}. MRI is, however, limited to imaging bed
particles that contain traces of water or oil, such as seeds.
Therefore we checked this important assertion more generally for
other particles, by freezing a water-saturated sample as explained
earlier. Figure \ref{overview} shows axial cuts through the bed for
three different density intruders. The results confirm the MRI data
for poppy seeds \cite{29}. At a density ratio $\rho / \rho_m \approx
0.5$, the intruder does not move with respect to its surroundings
and follows the convective flow. It has not separated from the first
black layer it was originally sitting on. The heavy intruder shows a
characteristic wake below it and is well above the first black
layer. Nearby particles are drawn under the sphere. The light
intruder is also clearly displaced from its original position but
does not exhibit the pronounced wake shown by the heavy intruder.
This indicates different mechanisms for the light and heavy intruder
by which they rise faster than convection. This behavior also holds
for the asymmetrically convecting bed. Intruders near $\rho / \rho_m
\approx 0.5$ follow the convective flow that drives them towards the
wall.

\begin{figure}
\begin{center}
\includegraphics[width=3.4in]{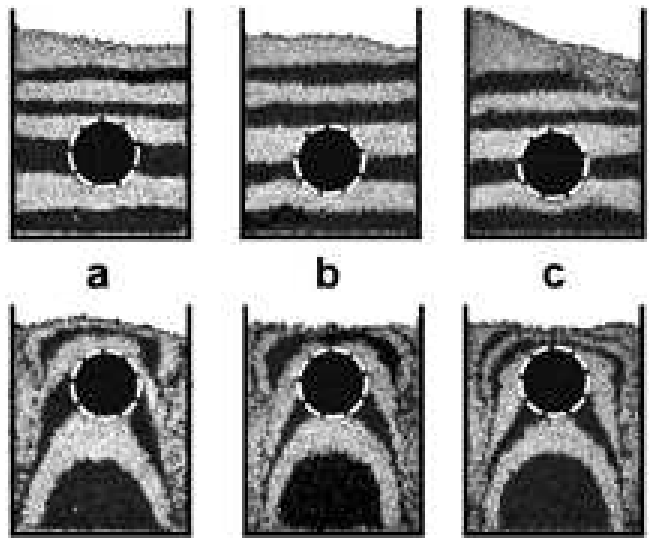}%
\caption{\label{mrivac} Absence of density dependence in vacuum. The
picture are MRI images of axial cuts through the bed for different
density intruders at $p=0.13$ kPa. System dimensions: $R=3.8$ cm,
$H=4.2$ cm and $D=1.6$ cm. The excitation parameters were $f=10$ Hz
and $\Gamma=3$. Upper row shows initial configuration. Lower row
shows system after $8\pm 1$ taps. Column (a) $\rho / \rho_m = 0.05$,
(b) $\rho / \rho_m = 0.45$, (c) $\rho / \rho_m = 2.4$.}
\end{center}
\end{figure}

{\it Effects of air and phase diagram}. MRI images of intruders
rising in an evacuated bed (Fig. \ref{mrivac}) show that the density
dependence of the rise time is due to the presence of air. Intruders
with different densities all rise with the surrounding bed. There is
no relative motion between the intruder and its vicinity. The only
transport mechanism is global convection \cite{23,41,29,4}.

\begin{figure}[t]
\begin{center}
\includegraphics[width=3.4in]{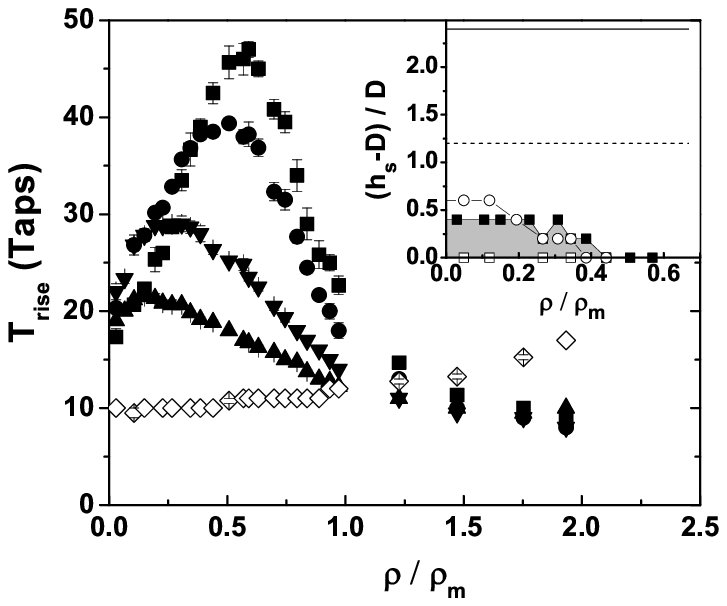}%
\caption{\label{roughpeaknphase} Rise time curves at different
pressures and the phase diagram in the rough cell. $\Gamma=5$,
$f=13$ Hz and $H=8.5$ cm. Intruder rise time $T_\textrm{rise}$
versus $\rho / \rho_{m}$ at different pressures $P$ in the rough
cell at $h_s=5.5$ cm: $(\blacksquare)$, $101$ kPa; ($\bullet$), $47$
kPa; $(\blacktriangledown)$, $13$ kPa; $(\blacktriangle)$, $6.7$
kPa; $(\lozenge)$, $0.13$ kPa. Inset: Phase diagrams delineating the
rising and sinking regimes for $d=0.5$ mm beds at various pressures
$P$. Above each phase boundary, intruders rise. Shaded area shows
the sinking regime at ambient pressure. Solid and dashed lines
indicate total bed height and starting height, respectively.
$(\blacksquare)$, $101$ kPa; $(\circ)$, $27$ kPa  and $(\square)$,
0.13 KPa.}
\end{center}
\end{figure}
\begin{figure}[t]
\begin{center}
\includegraphics[width=3.4in]{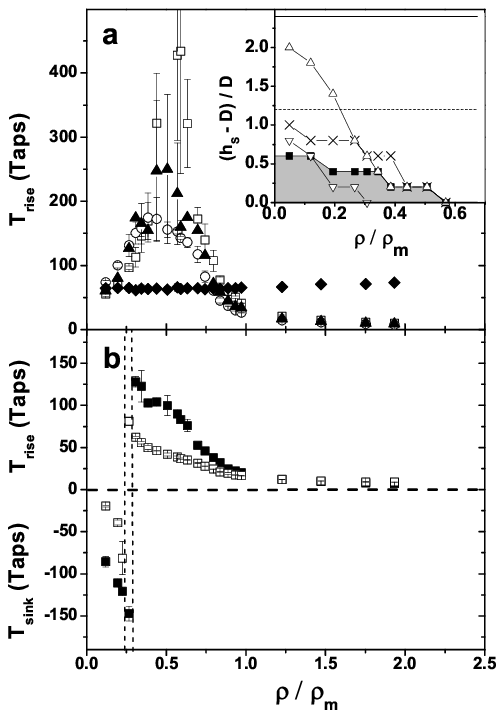}%
\caption{\label{smoothpeaknphase} Rise/sink time curves at different
pressures and the phase diagram in the smooth cell. $\Gamma=5$,
$f=13$ Hz and $H=8.5$ cm. (a) Intruder rise time $T_\textrm{rise}$
versus $\rho / \rho_{m}$ at different pressures $P$ in the smooth
cell at $h_s=5.5$ cm: $(\square)$, $101$ kPa; ($\blacktriangle$),
$40$ kPa; ($\circ$), $27$ kPa; $(\blacklozenge)$, $0.13$ kPa. Inset:
Phase diagrams delineating the rising and sinking regimes for
$d=0.5$ mm beds at various pressures $P$. Above each phase boundary,
intruders rise. Shaded area shows the sinking regime at ambient
pressure. Solid and dashed lines indicate total bed height and
starting height, respectively. $(\blacksquare)$, $101$ kPa;
($\times$), $27$ kPa; $(\vartriangle)$, $2.7$ kPa;
$(\triangledown)$, $0.67$ kPa in the smooth cell. (b) The divergent
sinking regime. $(\blacksquare)$, $13.3$ kPa with $\rho^*=0.28$ and
$(\square)$ $2.7$ kPa with $\rho^*=0.24$. Same conditions as in (a).
The vertical dashed lines correspond to $\rho^*$.}
\end{center}
\end{figure}
At atmospheric pressure, on the other hand, the air enables
intruders which are lighter or heavier than $\rho^{*}$ to rise
faster than convection (Fig. \ref{overview}). This result originates
from the interaction between the bed, the intruder and the
interstitial air which we will discuss later. However, this is not
the only possible outcome. The {\it same} system can exhibit sinking
behavior. This is one of our central results and is summarized in
Figs. \ref{roughpeaknphase} and \ref{smoothpeaknphase}. These data
show how pressure and initial vertical position in the bed determine
whether the intruder moves up or down.

In order to test the influence of convection we performed the same
experiments in rough and smooth cells, which differ in convection
speed and flow pattern, keeping all other parameters the same. In
both cases we used a $0.5$ mm glass medium. The rough cell exhibits
symmetric wall-driven convection. In the smooth cell convection is
significantly reduced and the flow is asymmetric as depicted in
figure \ref{convection}(b). In Figs. \ref{roughpeaknphase} and
\ref{smoothpeaknphase} the initial starting height is just above the
middle of the bed at $h_s=5.5$ cm. In both cases the curve either
peaks or diverges at $\rho^* / \rho_m \approx 0.5$. As the pressure
is lowered $\rho^* / \rho_m$ and the associated rise times decrease.
It is known from Kroll's work \cite{61} that, due to air drag, the
bed does not lift off as high as it would in vacuum. This has a
direct consequence on the wall-driven convective flow. There is less
shear with the side walls and thus, convection is reduced. As a
result intruders rise faster in vacuum than in the presence of air.
This leads to the shorter rise times we observe at low pressures. In
both cells the density at which the peak occurs decreases
significantly at low pressures \cite{29}.

We also observe a transition between a rising and sinking regime in
the smooth cell. Below a certain pressure intruders with densities
below $\rho^*$ sink. The resulting rise/sink time curve is shown in
Fig. \ref{smoothpeaknphase}(b). The data resemble those found
earlier by Yan et al. \cite{21}. When the pressure is lowered even
further the intruders rise again and the density dependence of
$T_\textrm{rise}$ vanishes.

This crossover between rising and sinking is not only controlled by
pressure but also by the vertical position of the intruder in the
bed. If the intruder is initially placed below a certain crossover
height $h_c$ it sinks. In general, $h_c$ depends on several other
parameters as well: The pressure $P$, the size of the medium $d$,
the density $\rho_m$, the total height $H$ and the wall roughness.

\begin{figure}
\begin{center}
\includegraphics[width=3.4in]{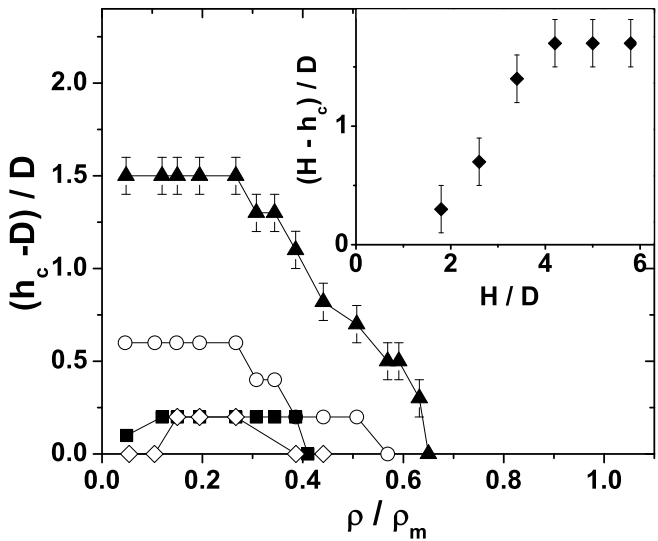}
\caption{\label{phaseplot} Phase diagram for different sized glass
media in the smooth cell at $P=101$ kPa. $\Gamma=5$, $f=13$ Hz and
$H=8.5$ cm. $(\blacktriangle)$, $0.25$ mm; $(\circ)$, $0.5$ mm;
$(\blacksquare)$, $1.0$ mm; $(\lozenge)$, $2.0$ mm. Inset: Crossover
height $h_c$ versus total height $H$ normalized by $D$ for $0.25$ mm
glass medium in the smooth cell.}
\end{center}
\end{figure}

In order to delineate the rising and sinking regimes, we map out
phase diagrams in terms of the normalized starting height,
$(h_s-D)/D$, and the normalized density, $\rho / \rho_m$ (insets to
Figs. \ref{roughpeaknphase} and \ref{smoothpeaknphase}, and main
panel of Fig. \ref{phaseplot}). The lines connecting the data in
these phase diagrams give the experimentally determined, normalized
crossover height $(h_c-D)/D$. Started above the lines, intruders
rise, started at or below the lines they sink. The grey areas in the
insets of Figs. \ref{roughpeaknphase}(a) and
\ref{smoothpeaknphase}(a) denote the sinking regimes at atmospheric
pressure. As the pressure is lowered $h_c$ initially increases and
then drops. The intruders, regardless of their density, rise from
the bottom of the cell ($h_c=D$) when $P<0.13$ kPa. In this regime
air effects do not play any role. In the rough cell, $h_c$ is lower
and the pressure dependence not as pronounced as in the smooth cell.
We believe that strong convection in the rough cell gives an upward
bias and therefore lower crossover heights compared to the smooth
cell. The difference between the rough and smooth cell might have
another cause. Increased shear at the walls could cause bed dilation
and therefore provide a shortcut for the air due to increased
permeability. This would lead to a decrease in air effects in the
rough cell.

The sinking of the intruder is caused by pressure gradients across
the bed exerting a net downward pull on the intruder \cite{29}.
Therefore, the phase diagram changes with the permeability of the
bed, since the magnitude of the pressure gradient is inversely
proportional to the permeability according to Darcy's law. We can
adjust the permeability $k$ by changing the size of the background
medium, since $k \propto d^2$. Figure \ref{phaseplot} shows the
phase diagram for glass beads of different sizes. As the medium size
increases, the sinking regime shrinks. This is expected since in the
limit of large permeability the pressure gradients are too weak to
induce sinking. Indeed, we find that above $d=2$ mm all intruders
rise with convection and air effects become negligible. The inset in
Fig. \ref{phaseplot} shows $(H-h_c)/D$ versus the normalized total
height $H/D$. The curve saturates as $H$ increases.

{\it Size dependence}. In the results shown so far, the relative
size between the intruder and the medium, $D/d$, has been kept at
$51$. It is vital for understanding the effect to see whether the
strong density dependence is still observed as $D/d$ approaches 1.
\begin{figure}
\begin{center}
\includegraphics[width=3.4in]{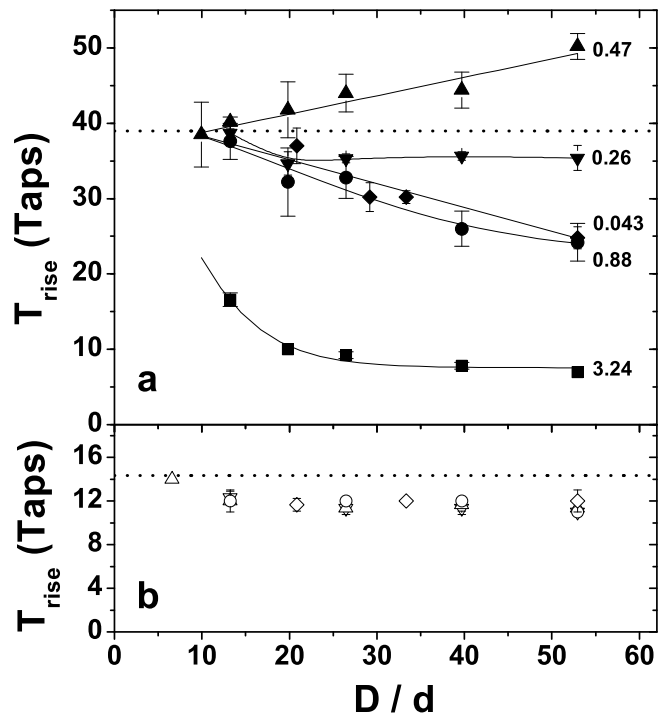}%
\caption{\label{diffsize} Intruder rise time $T_\textrm{rise}$
versus relative size $D/d$ for different densities at two different
pressures: (a) atmospheric pressure $P=101$ kPa and (b) low pressure
$P=0.13$ kPa. $\Gamma=5$, $f=13Hz$, $H=8.5$ cm and $h_s=4.0$ cm. The
numbers next to the line indicate the relative density. The dotted
line indicates the convective rise time \emph{without} the intruder.
Relative densities in panel b) are: $(\lozenge)$, $\rho /
\rho_m=0.043$; $(\triangledown)$, $\rho / \rho_m=0.26$;
$(\vartriangle)$, $\rho / \rho_m=0.47$; $(\circ)$, $\rho /
\rho_m=0.88$. The heaviest intruders ($\rho / \rho_m=3.24$) stop
rising roughly one intruder diameter below the surface.}
\end{center}
\end{figure}
Figure \ref{diffsize}(a) shows the size dependence of the rise time
for different intruder densities at atmospheric pressure. Size
affects the rise times notably. Both size and density dependencies
become less pronounced for smaller $D/d$. This might also explain
the results of Huerta et al. \cite{30}. They find the convective
rise time to be faster than most intruder rise times. However, they
measured convection in the absence of intruders which is different
from when an intruder is present. Fig. \ref{diffsize}(a) clearly
shows that convection rise time in the absence of an intruder
(dotted line) is lower than the rise time of the intruder at the
peak ($\rho / \rho_m = 0.47$). From our previous MRI measurements
\cite{29} and frozen-bed images (Fig. \ref{overview}) we know that
this intruder rise time reflects the actual convection time. We
believe that the large sphere blocks air flow, thereby decreasing
permeability which in turn slows down convection. At relative sizes
$D/d < 10$ the influence of the intruder on the bed diminishes and
rise time of the intruder at the peak approaches the convective rise
time without the intruder. Therefore, at small $D/d$ air-driven size
separation becomes negligible and convection dominates \cite{4}.

At pressures below $0.13$ kPa [Fig. \ref{diffsize}(b)], the rise
times are nearly independent of $\rho / \rho_m$ and $D/d$ and are
close to the convective rise time in the absence of an intruder. The
slight increase in $T_\textrm{rise}$ at low $D/d$ might be due to
geometric arching effects as discussed in \cite{46,9,34,35,36}.
Also, the heaviest intruder ($\rho / \rho_m = 3.24$) does not
surface at low pressures. It stops rising just below the surface and
appears to sink back into the fluidized layer. At atmospheric
pressure, the heavy intruders have a longer flight time than the bed
due to their larger inertia so that the bed is already condensed
when the sphere hits the surface. This explains why it does not sink
back in as much as it does at low pressure when the upper surface is
still fluidized at impact.

\begin{figure}
\begin{center}
\includegraphics[width=3.4in]{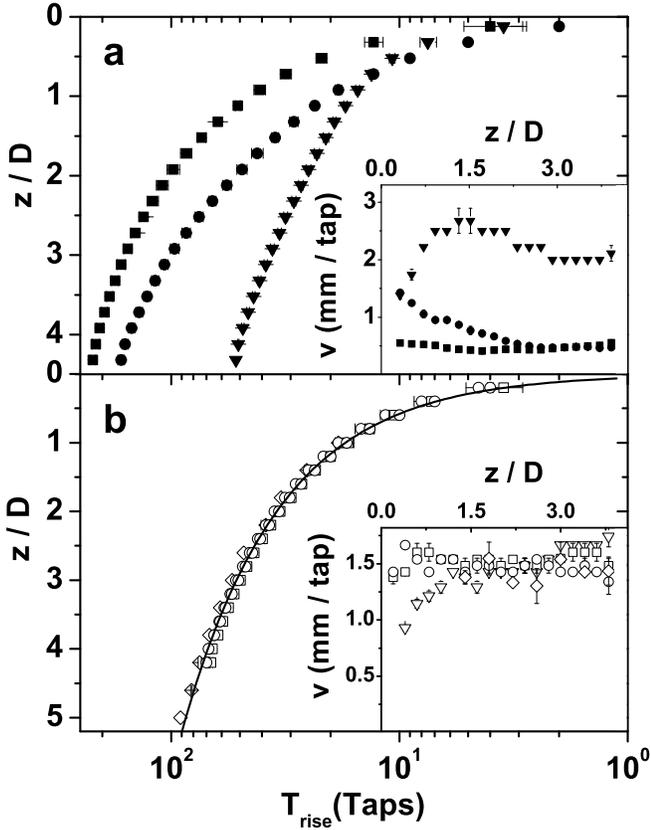}%
\caption{\label{depthpaper} Depth dependence of intruder rise time
in $0.5$ mm glass medium. Total bed height $H=13.5$ cm, $\Gamma=3.5$
and $f=13$ Hz. Upper panel (a): $T_\textrm{rise}$ versus $z/D$ for
different density intruders ($D/d = 50$) at atmospheric pressure.
$(\bullet)$, $\rho / \rho_m = 0.043$; $(\blacksquare)$, $\rho /
\rho_m = 0.52 \approx \rho^{*} / \rho_m$; $(\blacktriangledown)$,
$\rho / \rho_m = 3.3$. Inset shows the velocity at different depths.
Lower Panel (b): Same plot as above, but at $p=0.13$ kPa. $(\circ)$,
$\rho / \rho_m = 0.043$; $(\square)$, $\rho / \rho_m = 0.52 \approx
\rho^{*} / \rho_m$; $(\triangledown)$, $\rho / \rho_m = 3.2$;
$(\lozenge)$, convection without intruder. Solid line is a linear
fit with slope 0.058 tap$^{-1}$.}
\end{center}
\end{figure}

{\it Intruder trajectories}. In order to elucidate the mechanisms by
which light and heavy intruders rise faster than convection, we
tracked their positions during the ascent. Panel \ref{depthpaper}(a)
shows the rise time as a function of depth for different density
intruders at atmospheric pressure. The inset shows the displacement
per tap versus depth. Deep in the bed the light intruder moves as
fast as the intruder at the peak density $\rho^*$. It then speeds up
as it approaches the surface. The heavy intruder is always faster,
but slows down near the surface, because it falls back into the
fluidized top layer at the end of the cycle. Below $0.13$ kPa, the
curves collapse and the only transport mechanism is convection [Fig.
\ref{depthpaper}(b)]. This is consistent with the convective rise
time equation $z(t)=\xi \ln(1+T_\textrm{rise} / \tau)$ found by
Knight et al. \cite{37} given that in the present experiments the
bed height $H$ is comparable to the depth of the convection roll,
$\xi$. This implies that $T_\textrm{rise} \ll \tau$, leading to $z
\approx (\xi / \tau) T_\textrm{rise}$. The data in Fig.
\ref{depthpaper}(b) are well fit with $\xi/\tau=1.5$ mm/tap (solid
line).

\begin{figure}
\begin{center}
\includegraphics[width=3.4in]{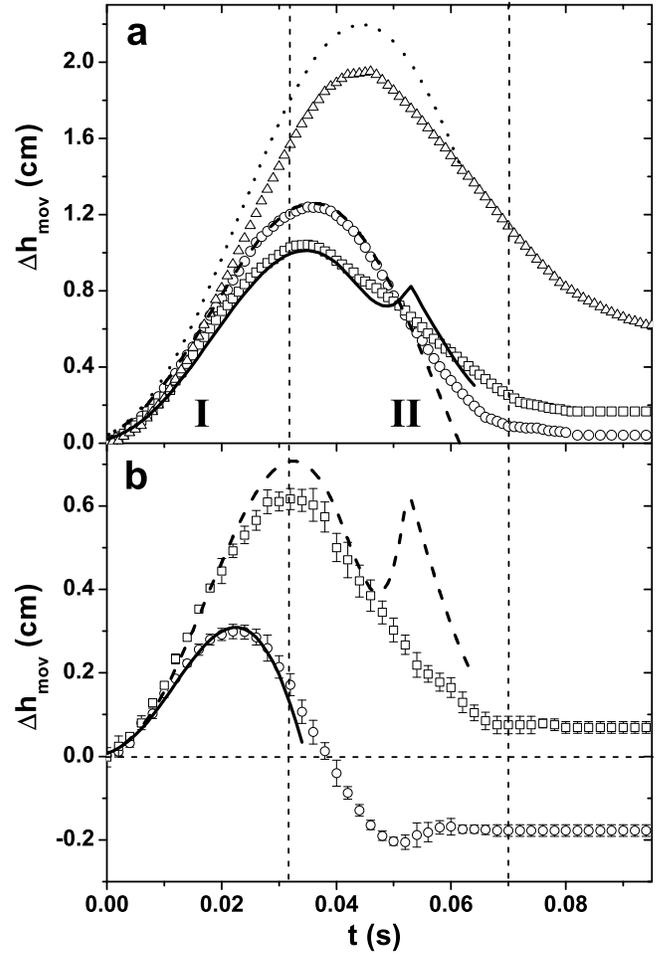}%
\caption{\label{trajectories} Intruder displacement in the shaker
frame. The vertical dashed lines delineate the part I and part II of
the cycle (see text). (a) Three different densities with $h_{s}=7.0$
cm in the rough cell. $(\square)$, $\rho / \rho_{m}=0.043$;
$(\circ)$, $\rho / \rho_{m}=0.52\simeq \rho^{*}/\rho_m$;
$(\vartriangle)$, $\rho / \rho_{m}=3.3$. The solid, dashed and
dotted lines are simulation results for trajectories of the light,
medium and heavy intruder (Section V). (b) Light intruder ($\rho /
\rho_{m}=0.043$) in smooth cell rising at $h_s=6.5$ cm $>h_c$,
($\square$), and sinking at $h_s=3.5$ cm $<h_c$, ($\circ$). The
solid and dashed lines are the respective simulation results. The
packing fraction was chosen to be $\phi=0.59$ and $\phi=0.63$ for
the rough and smooth cell, respectively. For the sinking intruder in
(b), the simulation was stopped at $t=36$ ms when the bed volume
underneath the intruder hits the bottom of the cell.}
\end{center}
\end{figure}

Using high-speed video to image the trajectory of the intruder
during one shaking cycle we arrive at a more detailed understanding
of the mechanism by which the intruders rise or sink. Figure
\ref{trajectories}(a) shows the trajectories of intruders for three
densities. The vertical lines delineate two parts of the shaking
cycle. In part I, air flows down to the gap that opens up at the
bottom the cell \cite{38,28}. In the second part, the gap starts to
close and air has to leave the bed. At $\rho^*$ the intruder has the
smallest net upward displacement. The heavy intruder rises higher
than all the others, as its inertia is large compared to air drag.
Therefore, its net upwards displacement happens in part I. The light
intruder on the other hand does not rise higher than the one at
$\rho^*$. However, it slows down in part II, leading to a higher
{\it net upward} displacement than the intruder at $\rho^*$.

\begin{figure}
\begin{center}
\includegraphics[width=3.4in]{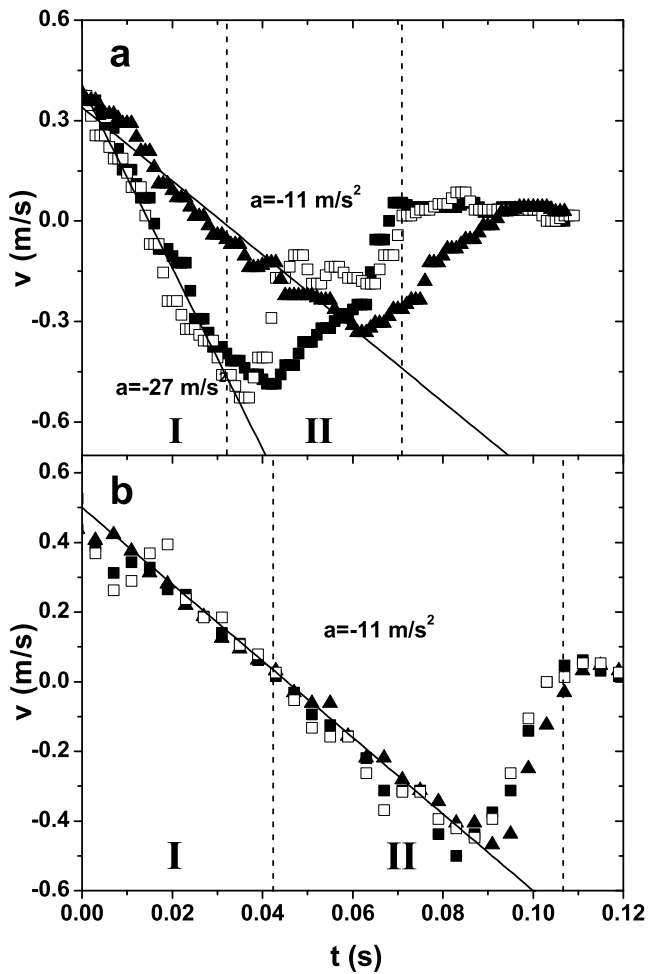}%
\caption{\label{velocities} Intruder velocity in the lab frame at:
(a) atmospheric pressure $P=101$ kPa and (b) low pressure $P=0.13$
kPa. The vertical dashed lines delineate the part I and part II of
the period (see text). Three different densities with $h_{s}=6.0$ cm
in the rough cell. $(\square)$, $\rho / \rho_{m}=0.043$;
$(\blacksquare)$, $\rho / \rho_{m}=0.52\simeq \rho^{*}/\rho_m$;
$(\blacktriangle)$, $\rho / \rho_{m}=3.3$. The accelerations denote
the the slopes of the corresponding solid lines.}
\end{center}
\end{figure}

The velocity trajectories of the three intruders in the lab frame
[Fig. \ref{velocities}(a)] show that the heavy intruder experiences
little retardation. The slope of the velocity curve is close to $-g$
as denoted by the solid line throughout part I and II. It only slows
down once the bed starts to condense after part II. Conversely, the
lighter intruders are subject to a significant drag force as evident
from their high downward acceleration: the first part of the
velocity curves is well fit by the $-3g$ slope of the solid black
line. They both slow down at the beginning of part II when the air
flow reverses.

Figure \ref{trajectories}(b) shows the trajectories for light
intruders ($\rho \ll \rho^{*}$) above and below the crossover
height. Both intruders are slowed down in part II by the escaping
air. In part I, however, the sinking intruder does not rise as high
as the ascending intruder which causes a net downward displacement
after part II. This observation is crucial for understanding the
underlying mechanism.

\begin{figure}
\begin{center}
\includegraphics[width=3.4in]{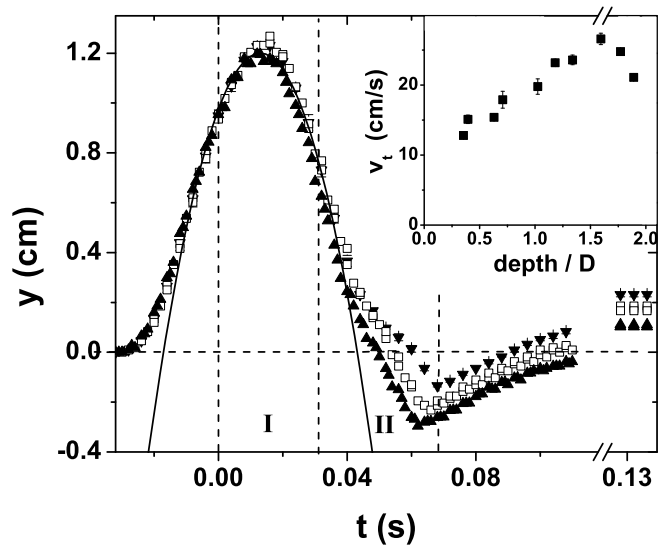}%
\caption{\label{kinkdepth} Trajectory of light intruder ($\rho /
\rho_{m}=0.043$) in the lab frame in $d=0.5$ mm glass medium at
different depths in the rough cell. $\Gamma=5$, $f=13$ Hz.
$(\blacktriangle)$, $h_s=4.0$ cm; $(\square)$, $h_s=5.5$ cm;
$(\blacktriangledown)$, $h_s=7.5$ cm. The black line is a parabola
with acceleration $a=-27$ $\textrm{m/s}^2$. Inset: $(\blacksquare)$,
terminal velocity versus depth.}
\end{center}
\end{figure}

Several studies \cite{30,44,16,6} have suggested that the right hand
side of the peak is caused by inertia. In this model heavy intruders
penetrate more easily through the bed which is held back by shear
with the walls. In vacuum, where gravity and wall-shear are the only
forces acting on the bed, our data for the intruder trajectories
show no evidence of this. All intruders experience the same
acceleration close to $g$ during flight [Fig. \ref{velocities}(b)].
Therefore, the wall shear transmitted through the bed exerts only a
small force on the intruder in the absence of air with little or no
density dependence.

Figure \ref{kinkdepth} shows the light intruder trajectory for
different depths in the lab frame. As in Fig. \ref{trajectories},
the light intruder slows down in part II of the cycle and reaches a
terminal velocity. There is a pronounced kink at the onset of the
slowdown. The closer the intruder is to the surface the more it
slows down, which increases the upward displacement and is in
agreement with our previous observation in Fig. \ref{depthpaper}.
The inset shows how the terminal velocity decreases as the intruder
nears the surface.

{\it Air drag}. In the following we look at the influence of the
pressure gradients inside the bed on the motion of the intruder.
These pressure gradients can be significant in vibrated beds
\cite{38}. To see its effect on the intruder motion we performed a
simple experiment: Two intruders, one with $\rho \approx \rho^{*}$
and the other with $\rho \gg \rho^{*}$, were put on top of a bed.
After tapping, the trajectories of the two intruders differ notably
at atmospheric pressure. While the heavy intruder is visible
throughout the cycle, the lighter one is ``sucked'' into the bed and
disappears. At the end of the shaking cycle, the light intruder
reappears, while the heavy one burrows into the bed due to its large
inertia \footnote{The high-speed movies demonstrating the drag on
the intruders can be found on
http://jfi.uchicago.edu/$\sim$jaeger/granular2/videos/ Brazil Nut
videos.html}. In vacuum, the air drag vanishes and they rise to the
same height. No relative motion is observed between the two
intruders (except that at the end of the cycle the heavy intruder
dives into the bed again). To see if Stokes drag plays any
appreciable role, the intruders were put on a plate and subsequently
tapped. They rose to the same height within our experimental
resolution demonstrating that Stokes drag on the intruder is
negligible.

Summarizing, we find that the drag on the intruder inside the bed is
substantial and originates from the pressure gradient set up by the
bed during the shaking cycle. In the absence of the bed, where
Stokes drag is applicable, the air drag has no measurable effect on
the intruder.

Huerta et al. \cite{30} proposed a different model for the formation
of the peak in the rise time versus $\rho / \rho_m$. They suggest
that the peak results from a competition between convection and
inertia effects, both of which should also occur in vacuum. However,
in our experiments the density dependence always vanished at low
pressures. We checked this result explicitly for $2$ mm Tapioca
pearls, a medium that is similar to the $3$ mm pearls used by Huerta
et al. The peak observed at atmospheric pressure flattened out in
vacuum. The reason why air is still important for these relatively
large and porous media is that Tapioca is a very light material
($\rho_m = 1.2$ g/ml) and much more susceptible to drag forces than
glass beads ($\rho_m = 2.5$ g/ml) of comparable size.

The literature on granular size separation mainly deals with two
systems: The single intruder on one hand and binary mixtures which
separate into two phases on the other hand. In the former system the
interaction is only between the intruder and the bed, while in
binary mixtures the interactions between the particles that make up
one species need to be taken into account as well. In a model
proposed by Hong and coworkers \cite{7} the separation of binary
mixtures depends on the granular temperatures of each species. The
species with the larger diameter condenses at the bottom of the
container - the so-called reverse Brazil nut effect - if $D / d
\geqq \rho_d / \rho_D$, where $\rho_d$ and $\rho_D$ are the
densities of the small and large species, respectively. This model
was shown to be in agreement with a recent experiment on separating
binary mixtures \cite{25}  with particle diameters in the range
$2-22$ mm.

In order to test for the onset of interactions between multiple
intruders in our system we investigated cases of up to nine
intruders in a variety of initial geometrical arrangements. All
intruders were placed at the same starting height in a large, rough
cell ($R=6.0$ cm) with glass beads as bed material. We used $d=0.5$
and $1.0$ mm beads and intruder sizes ranging from $D/d=12.5$ to
$51$, and density ratios $\rho_d / \rho_D$ between $0.3$ and $2$.
Therefore the condition derived by Hong et al. for the reverse
Brazil nut effect to occur was always satisfied. However, none of
the configurations sank. Moreover, the intruders moved as a compound
with rise times that did not show any appreciable difference to our
single intruder experiment. This held independently of the initial
spatial arrangement of multiple intruders. This result has two
implications: If the model proposed by Hong et al. \cite{7} were
applicable, then many more intruders are needed to establish a
granular temperature that gives rise to condensation at the bottom.
Secondly, in the regime of our experiment the size ratio $D/d$ is
not the only relevant parameter and air effects clearly cannot be
neglected.

\section{IV. Model}

The experimentally established connection between the density
dependent rise times and the presence of pressure gradients during
shaking can be understood using a simple model that we develop in
this section. This model takes into account the interactions between
the bed, the intruder and the air flow during each shaking cycle and
enables us to deduce the net displacement of the intruder after each
tap. In the next section we will incorporate the model into a
simulation to provide a more quantitative comparison to our
experimental results.

Each shaking cycle is divided into two parts: During Part I, the bed
lifts off the bottom of the cell and a gap beneath opens up. The
low-pressure region formed at the bottom causes air to flow down. As
a result the bed and the intruder experience a downward force.
Smaller media with lower permeability increase that air drag. The
second part of the cycle starts when the gap starts to close again,
the pressure gradient reverses sign and air flows upwards until the
bed hits the bottom of the cell \cite{38,28,17}. Let us now make
some simplifying assumptions: We treat the bed as a porous solid
with a constant packing fraction $\phi$; Any gap caused by relative
motion between intruder and bed is immediately filled up; The
pressure gradients created during the shaking cycle do not depend on
the position within the bed; Horizontal pressure dependence due to
the presence of walls is negligibly small; Finally, we do not
consider convection.

Within this model the vertical pressure gradient across the bed is
governed by Darcy's law
\begin{equation}\label{darcy}
\frac{\partial P}{\partial z} = \frac{\mu}{k} u,
\end{equation}
where $k$ is the permeability of the bed, $\mu$ the viscosity of the
interstitial fluid and $u$ the fluid velocity outside the bed. The
pressure $P(z,t)$ is a function of time and the vertical coordinate
$z$ only. Using Eq.(\ref{darcy}) we find the drag force on the bed
medium $F_m$ and intruder $F_{int}$:
\begin{equation}\label{Fdb}
F_{m} =  \partial P / \partial z \cdot V_{m} = \frac{\partial P /
\partial z}{\phi\cdot \rho_{m}} m_{bed}
\end{equation}
\begin{equation}\label{Fint}
F_{int} =  \oint_S P \hat{z} \vec{dS} =
\partial P / \partial z \cdot V_{int} = \frac{\partial P /
\partial z}{\rho} m_{int}
\end{equation}
In the last integral we used the assumption that $\partial P /
\partial z$ is a function of time only. This drag term, $F_{int}$,
has been used before to describe the effect of air on a particle in
a vibrated bed \cite{17,27,45}. Comparing $F_m$ and $F_{int}$, it is
clear that the bed and the intruder experience the same acceleration
when $\rho / \rho_m = \phi$.

We now compare this drag force and the viscous drag acting on the
intruder. The viscous drag in laminar flow is given by Stokes
formula:
\begin{equation}
F_{\textrm{Stokes}}=6 \pi \eta r u
\end{equation}
In our setup, $r=0.0125$ m, $u\approx A(2\pi f)=0.6$ m/s and
$m_{int}$ ranges from $0.5$ to $66.7$ g, so $F_{\textrm{Stokes}} /
(m_{int}\cdot g)$ ranges from $5.2\cdot 10^{-4}$ to $3.9 \cdot
10^{-6}$. Therefore, we can safely ignore simple Stokes drag on the
intruder. The viscous drag on the bed has been accounted for in
Darcy's law, where the bed is treated as a continuous porous block.
The intruder is also subject to the force due to pressure gradients
as shown in equation (\ref{Fint}). The typical gradients in our bed
with $0.5$ mm glass bead during the shaking cycle can be estimated
from Darcy's law (\ref{darcy}): $\partial P /\partial z \approx 25
\cdot 10^{3}$ Pa/m using $\mu=1.8 \cdot 10^{-5}$ Pa s for air and
the Carman-Kozeny relation for the permeability of a random packed
bed of spheres \cite{60}
\begin{equation}
k=\frac{d^2 (1-\phi)^3}{180 \phi^2}.
\end{equation}
Therefore $F_{int} / (m_{int}\cdot g)$ is between $0.31$ to $42$.
Even though Stokes drag does not influence the intruder appreciably,
the forces due to pressure gradients are substantial.

It is useful to estimate the effective acceleration acting on the
bed in part I. The apex of an object's trajectory on which a
constant drag force $F_d$ acts and an initial velocity $v_0$ is
given by $h_{max}=v_0^2 / (2g_{eff})$, where
\begin{equation}\label{geff}
g_{eff}=g\left(1+\frac{F_d}{mg}\right)=g\left(1+\frac{\mu u}{k \phi
\rho_m g}\right) .
\end{equation}
after substituting (\ref{Fdb}).

Substituting $\rho_m=2500$ kg/$m^3$ for glass, a typical packing
fraction $\phi=0.54$ and $u=0.6$ m/s we obtain $g_{eff}=27$ m/$s^2$,
close to the acceleration of the light intruder in Figs.
\ref{velocities}(a) and \ref{kinkdepth} until it slows down and the
kink develops. We expect the intruder at density $\rho^*$ to have
the same acceleration as well in part I, since it moves with the
bed. This is indeed the case as shown in Fig. \ref{velocities}(a).
The intruder has the same acceleration before it slows down in part
II. The slow down is not as abrupt as for the light intruder.

At density $\rho^* / \rho_m = \phi$ the accelerations of the
intruder and the bed are the same, so there will be no relative
motion. In order to illustrate the consequences of air drag on
different density intruders it is instructive to look at the two
parts of the cycle for an extremely light and heavy intruder.

Part I: If $\rho \gg \rho^{*}$, the downward acceleration on the bed
is greater than for the intruder as shown by Naylor et al.
\cite{26}. The intruder will push the bed above it. The closer the
intruder is to the surface the less material it has to displace. If
$\rho \ll \rho^{*}$, the intruder will be pushed against the bed
below it. Two things can now happen: If the intruder is near the
bottom of the bed, less material has to be displaced and the
intruder can sink with respect to the bed. If the intruder is
located close to the surface, the material below will block the
downward motion of the intruder and it moves with the bed.

Part II: For intruders with $\rho \gg \rho^{*}$ the time of flight
is longer than for the bed. While the bed collapses the intruder is
still in the upward motion with respect to the bed. The escaping air
has little effect on the intruder and the void below it is filled up
with bed material. By the time the trajectory of the intruder has
passed its apex the bed has almost collapsed and it will have a soft
landing on the bed resulting in a net upward displacement. The light
intruder with $\rho \ll \rho^{*}$ will be accelerated upwards with
respect to the bed due to its low inertia. If it is near the bottom,
it cannot displace much of the material above it. Above the
crossover height, however, it has enough inertia to displace the
material above and therefore will rise with respect to the bed.

The net displacement of the intruder after one shaking cycle is the
sum of displacements in part I and II. Therefore, intruders with
densities different than $\rho^*$ rise faster than convection or,
for the case of light intruders started initially below a crossover
height, will sink. In summary our model makes the following
predictions:
\begin{itemize}
\item At density $\rho^* / \rho_m = \phi$ the intruder does
not move with respect to the bed. In our simplified model without
convection it would stay put. In reality, however, there is always
some convection and the intruder will follow whatever background
flow there is in the system. If the flow is symmetric wall-driven
convection, then the intruder will surface in the middle of the
cylinder. In the case of the asymmetric convection, the intruder
will be driven to the side of the container.
\item At all the other densities, the intruder will
experience relative motion with respect to the bed due to air drag.
The intruder will push the bed above or below, depending on the
relative accelerations during the cycle. The dynamics between the
intruder and the bed caused by air drag enable light and heavy
intruders to move faster through the bed than those at $\rho^*$.
\item There exists a crossover height below which light intruders
with $\rho < \rho^*$ sink.
\item The density dependence should vanish when
the relative size $D/d$ becomes small. Even though the acceleration
due to the drag force on the intruder $F_{int}/m_{int}$ is
independent of $D$, the absolute value of its inertia decreases with
$D^3$. Therefore smaller intruders displace less bed than larger
ones with the same acceleration. Intruders with small size ratio
$D/d$ rise with convection as is evident in Fig. \ref{diffsize}(a).
This is also the reason why Knight et al. \cite{4} did not observe
any deviation from convection when they measured intruder rise
times. All their intruders had a relative size $D/d\leq 10$.
\item The validity of this model is independent of the phase of
the sine wave excitation. In general, this is true for any
excitation in which the acceleration exceeds $g$ and therefore
causes the bed to lift off. Whenever a gap opens up, the air flows
downward, independent of whether the cell moves up or down at that
point. Consequently, any excitation can be delineated into two
phases in which the air flows down first and then up.
\end{itemize}
A consequence of these considerations is the existence of a
crossover height. From symmetry arguments one would naively expect
it to be in the middle of the bed. However, convection and
fluidization of the upper layers cause an upwards bias in the
intruder motion. Moreover, the inertia of the intruder breaks the
symmetry. Despite the total impulse during the cycle due to drag
being close to zero, there will still be a net movement. Even if the
intruder did not collide with the bed during flight it would not
return to its initial position. This is due to the different initial
velocities at the beginning of part I and part II of the cycle.
These factors explain why the crossover height is usually not in the
middle of the bed.

Despite explaining main features of the separation mechanism the
model fails to account for some of our observations. One is the
decrease of $\rho^*$ with decreasing pressure (Figs.
\ref{roughpeaknphase} and \ref{smoothpeaknphase}). Even though
$\rho^*/\rho_m \lesssim \phi$, it is unlikely that $\phi$ would
decrease to below $0.4$. One possible explanation is that the
pressure gradients are not spatially uniform at lower pressures. In
this case equation (\ref{Fint}) would no longer hold.

Another feature of the data the model does not capture is the
non-monotonic behavior of the crossover height $h_c$ as a function
of pressure (insets of Figs. \ref{roughpeaknphase} and
\ref{smoothpeaknphase}).

\section{V. Simulation}
In order to test key aspects of our model we performed a numerical
simulation. In particular the computation of the pressure
distribution in the bed during one shaking cycle sheds light on one
main assumption in the model: the linear change of pressure with
depth at all times.

\begin{figure}
\begin{center}
\includegraphics[width=2.0in]{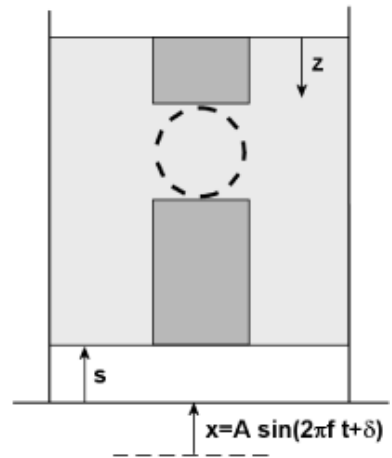}%
\caption{\label{sketch} Sketch of the vibrating bed indicating the
variables used in the model. The dark areas above and below the
intruder are the cylindrical bed volumes displaced by the intruder
during the shaking cycle. $s$ is the height of the gap between the
lower edge of the bed and the cell bottom.}
\end{center}
\end{figure}

The starting point is the Gutman model \cite{40,38} for the motion
of a vibrated porous bed. It essentially models a porous piston in a
vibrated cell subject to drag forces as the air pulsates through the
bed. There is no friction between the bed and the walls and the air
flow is assumed to be governed by Darcy's law (\ref{darcy}), as
before. The model also allows for the isothermal compression of the
gas \cite{40}. Figure \ref{sketch} shows a schematic view of the
system and introduces the relevant variables.

The pressure $P(z,t)$ obeys the diffusion equation \cite{60}
\begin{equation}\label{diffusion}
\frac{\partial P}{\partial t} = \frac{P_0 k}{\mu(1-\phi)}
\frac{\partial^2 P}{\partial z^2}.
\end{equation}
This equation together with Eq. (\ref{darcy}) are subject to the
following boundary conditions: At the top of the bed at $z=0$, the
pressure $P=P_0$, where $P_0$ is the ambient pressure. At $z=H$, the
continuity equation holds and
\begin{equation}
\frac{d(\rho_{gas}s)}{d t}=\rho_{gas,0} u=\left.
-\rho_{gas,0}\frac{k}{\mu}\frac{\partial P}{\partial z}\right
|_{z=H},
\end{equation}
where $\rho_{gas,0}$ is the density of the gas at atmospheric
pressure $P_0$. Since $\rho_{gas} \approx \rho_{gas,0}$ and $s
(d\rho / dt) \ll \rho_{gas,0} (ds / dt)$ this reduces to
\begin{equation}\label{boundary}
\frac{ds}{dt}=\left. -\frac{k}{\mu}\frac{\partial P}{\partial
z}\right |_{z=H}.
\end{equation}
Note that this approximation implies that the pressure variations
are small compared to the absolute pressure. In our system, the
pressure varies by $\approx 4$ kPa. Therefore, we require $P_0 \gg
4$ kPa.

The equation of the bed during the flight is given by
\begin{equation}\label{eom}
\frac{d^2(s+x)}{dt^2}=-g+\frac{(P|_{z=H}-P_0)}{\rho_m \phi H},
\end{equation}
where $x(t)=A \sin(2\pi f t + \delta)$ describes the oscillatory
motion of the cell in the lab frame. The bed lifts off when the
acceleration of the cell reaches $-g$. Thus,
$\delta=\arcsin(1/\Gamma)$. We integrate the diffusion equation
(\ref{diffusion}) numerically subject to the boundary conditions
using the Crank-Nicholson scheme \cite{39}.

The first step in the simulation establishes the pressure
distribution inside the bed during a shaking cycle as a function of
depth and time. Also the equation of motion of the whole bed
(\ref{eom}) is computed. We neglect any pressure redistribution due
to the presence of the intruder. This would lead to small
corrections that depend on the intruder geometry.

The second step determines the rise and sink time of different
density intruders at a specified starting height $h_s$. The
simulation integrates the equation of motion of three masses: The
bed volumes above and below the intruder and the intruder itself
(Fig. \ref{sketch}). These three objects ($i=1,2,3$) are subject to
drag forces and their displacement $y_i$ in the shaker frame obeys
\begin{equation}\label{drag}
\frac{d^2(s+y_i)}{dt^2}=-g+\frac{\oint_{S_i}P \hat{z}
\vec{dS}}{\rho_i V_i}.
\end{equation}

The simulation computes the net displacement after each tap, which
is the difference between initial and final position of the intruder
plus a small, fixed upwards displacement to account for convection.
The intruder is allowed to displace the upper and lower bed volumes
during flight, and after the shaking cycle any gaps are filled up to
freeze the intruder in place.

In order for the intruder to sink, there must be a net downward
displacement from its original position at the end of each shaking
cycle: The intruder must either penetrate the material below it or,
more likely, some of this material must be displaced sideways upon
hitting the bottom of the cell. In order to avoid introducing more
parameters and assumptions to the simulation for modeling the bed
displacement we stop the simulation at this point. Nevertheless, we
suggest that the region inaccessible to the simulation is where
sinking is possible according to our model.
\begin{figure}
\begin{center}
\includegraphics[width=3.4in]{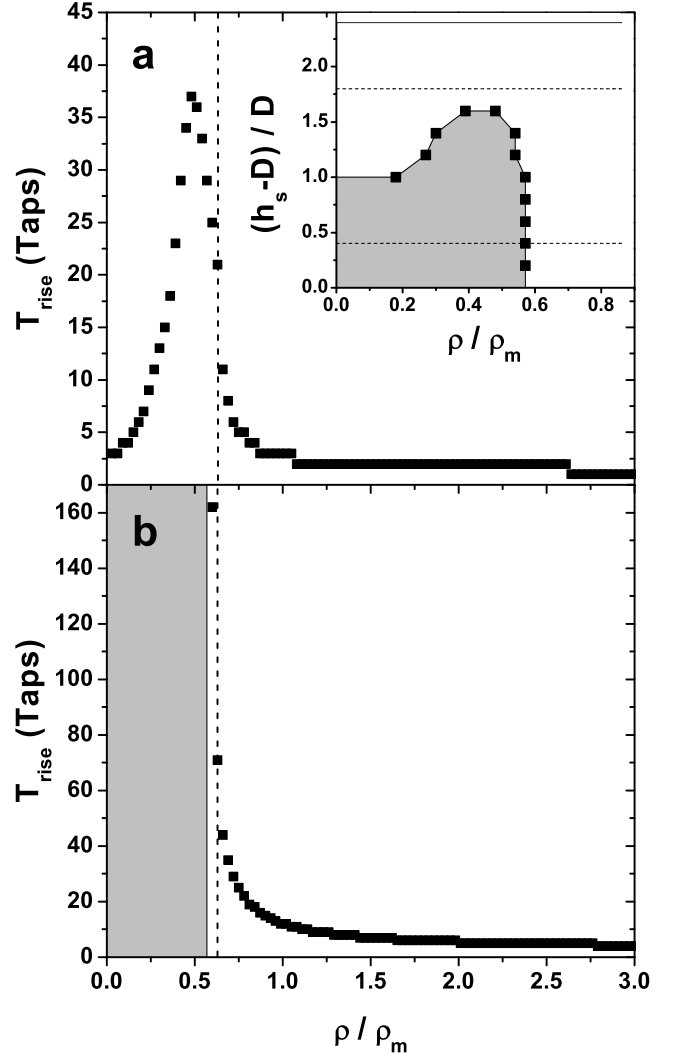}
\caption{\label{simulation} Simulation results of rise time
$T_\textrm{rise}$ versus relative density $\rho / \rho_m$ at two
starting heights above and below the crossover height: (a) $h_s=7.0$
cm and (b) $h_s=3.5$ cm. The parameters are: $H=8.5$ cm, $f=13$ Hz,
$\Gamma=5$, $\phi=0.63$ (vertical dotted line) and the convection
displacement is $0.7$ cm / tap. The inset in (a) shows the phase
diagram of this system at atmospheric pressure. Solid and dashed
lines indicate total bed height and the two starting heights in (a)
and (b), respectively. The gray area indicates the region
inaccessible to the simulation (see text).}
\end{center}
\end{figure}

Figure \ref{simulation} shows the results of the simulation for our
standard system: $H=8.5$ cm, $\Gamma=5$, $f=13$ Hz, $d=0.5$ mm,
$D=2.5$ cm, $P=101$ kPa and the convection displacement is $0.7$
mm/tap. The latter can be obtained from the experiments as follows.
From Fig. \ref{roughpeaknphase} we know that at $\rho^*$ the
intruder traverses $35$ mm in $45$ taps. Since the convection speed
does not change appreciably with depth (Fig. \ref{depthpaper}), the
displacement due to convection is $\approx 35/45=0.7$ mm/tap. The
packing fraction was chosen to be close to the random closed pack
limit, $\phi=0.63$, but any value between $0.5-0.63$ yields similar
results.

Comparing the simulation (Fig. \ref{simulation}) to the experimental
data from Figs. \ref{roughpeaknphase}, \ref{smoothpeaknphase} yields
quantitative agreement of key features: Both the peak and the
divergence in Fig. \ref{simulation} occur at $\rho^* / \rho_m = 0.5$
and $0.6$, respectively. Even though sink times cannot be obtained
in a quantitative fashion from the simulation without introducing
additional parameters, the grey area in Fig. \ref{simulation}(b) is
consistent with the sinking regime from the experiment.

Moreover, the simulated intruder trajectories closely match the
high-speed measurements in Fig. \ref{trajectories}(a): Denser
intruders rise higher and the apex occurs at later times. Light
intruders on the other hand, slow down in part II of the cycle. This
slow-down is responsible for the net displacement of the light
intruder with respect to the surrounding bed. The trajectories of
the light intruders cross those at $\rho^*$ at approximately
$t=0.05$ s, which agrees well with our simulation result. The kink
in the simulated light intruder trajectory is caused by the intruder
traversing the gap created above it in part I. In the experiment,
this gap is likely to get partially filled thereby impeding the
intruder motion. The simulated trajectories of the light intruder
above and below the crossover height [Fig. \ref{trajectories}(b)]
agree well with the data apart from the kink due to the presence of
the gap. The trajectory of the sinking intruder is only simulated
until the lower bed volume hits the bottom of the cell.

The simulation of a shaking cycle stops when the entire bed hits the
bottom of the cell. In the experiments, however, the intruder keeps
moving past that point as shown in Fig. \ref{trajectories}. This is
because the bed dilates during flight and condenses as the bottom
layers hit the cell bottom. This condensation takes approximately
$0.02$ s. The heavy intruder keeps moving even longer, since it is
penetrating the bed due to its large inertia. These effects are not
considered in our simulation and, given our results, they seem to be
second order effects.

The top boundary of the grey sinking regime is not flat [inset Fig.
\ref{simulation}(a)]. Instead it dips down at low densities.
Experimentally, however, we do not observe this for the $0.5$ mm
glass medium. Nevertheless, Fig. \ref{phaseplot} suggests that this
dip may exist for large permeabilities.

Due to the restrictions on $P_0$ by virtue of the approximations
made in equation (\ref{boundary}), we cannot probe very low
pressures in the simulation. Down to $P=40$ kPa $\rho^*/\rho_m$ only
shifted to $0.45$, which is a smaller reduction than observed in the
experiment \cite{29}. Even though the pressure change does not seem
to affect $\rho^*$ significantly, the crossover height goes up at
lower pressures. Moreover, making convection stronger by choosing a
higher value for the convection displacement in the simulation
decreased the crossover height.

Summarizing our simulation results, we find that important
parameters, such as peak and divergence position agree well with
experiments. The reason why the peak does not occur exactly at
$\phi$ is due to the fact that the pressure is not changing linearly
with depth at all times. This is a key assumption in our model. The
simulation shows departures from this assumption. The intruder
trajectories are in good quantitative agreement with the
experimental data (Fig. \ref{trajectories}) within the limits of our
simulation. We do not expect a full quantitative correspondence with
the experiment due to several factors the simulation lacks. It does
not account for penetration of the intruder through the bed, it
lacks any fluidization during flight and the experimental excitation
is not perfectly sinusoidal as assumed in the simulation. Moreover,
the bed masses that are displaced are assumed to be columns, which
is another idealization. Nevertheless, given the simplifying
assumptions, our model captures the essential features of the Brazil
Nut phenomenon and qualitatively reproduces the experimentally
observed phase diagrams in Figs. \ref{roughpeaknphase} and
\ref{smoothpeaknphase}.

A non-monotonic rise time versus density curve that peaks at $\rho^*
/ \rho_m \approx \phi$ has previously also been seen in a model
proposed by Rhodes et al \cite{27}. These authors use an equation of
motion for the intruder that includes three additional forces: two
types of bed retardation and an added mass term. It requires three
fitting parameters that are not directly obtainable from experiment.
Our model on the other hand needs only one directly-measurable
parameter, the convection displacement, which has a clear physical
origin.
\section{VI. Conclusions}
The role of air in vibration-induced granular size separation can be
dramatic. In media which are sufficiently small and light, air
effects dominate the process of size separation and can cause
intruders to sink that would rise otherwise. This process is
essentially decoupled from background convection. MRI, freezing the
bed with water, and high speed video measurements enabled us to
study the flows inside the bed and elucidate the separation
mechanism.

A key finding is that even though Stokes drag on the intruder in the
absence of the bed is negligible, the drag caused by bed-induced
pressure gradients is not. For one-inch diameter intruders in
half-millimeter bed material the gradients can lead to forces
several tens of times the intruders own weight. These forces cause
relative motion with respect to the bed during one shaking cycle.
This leads to a phase diagram that separates rising and sinking
behavior as a function of depth and pressure. Intruders with
$\rho/\rho_m < \phi$ sink if started below some crossover height.
Our model gives a unified description of the rising and sinking of
an intruder that agrees with previous experimental results.

The air drag also causes a size and density dependence of the speed
with which intruders rise or sink. We find that at $\rho^{*} /
\rho_m \approx \phi$ the intruder moves with the convective
background. Heavier intruders rise faster than convection, while
lighter intruders rise or sink with respect to convection depending
on their initial vertical position in the bed. This result as well
as measurements of the flight trajectories of the intruder during a
tap are in quantitative agreement with the model we developed.

In this model, heavy intruders rise  because bed particles are held
back more strongly by drag due to pressure gradients. Light
intruders, on the other hand, are more easily buffeted by the air
currents flowing through the bed; if placed initially near the
bottom, this can push the intruder downward, whereas if placed near
the top the intruder can get driven toward the upper surface.
However, we cannot explain the decrease of $\rho^*$ as the pressure
is lowered in the framework of our model. This might be due to
simplifying assumptions, such as constant packing fraction, lack of
fluidization and intruder penetration. Despite the fact that our
model cannot account for all the rich details this experiment
yields, it does unify previously disjointed pieces of data, mainly
the rising and sinking of light intruders.

The air driven size separation is a non-equilibrium phenomenon. The
different dynamics during the two parts of the shaking cycle explain
the observed results. Even though our studies concentrated on the
single intruder situation, similar considerations can be made for
multi intruders and binary mixtures where the size difference is
large and the bed permeability is small.

\begin{acknowledgements}
\section{Acknowledgements}
We thank N. Mueggenburg and E. Corwin for fruitful discussions and
J. Rivers for help with the MRI. P.E. acknowledges the hospitality
of the James Franck Institute during his stay. This work was
supported by the NSF under CTS-0405619, and under MRSEC, DMR-0213745
and by DOE under W-7405-ENG-82. MEM acknowledges support from the
Burroughs-Wellcome Fellowship No. 1001774.
\end{acknowledgements}

\end{document}